\documentstyle[12pt]{article}
\newcommand{\be}[1]{\begin{equation} \label{(#1)}}
\newcommand{\ee}{\end{equation}}
\newcommand{\ba}[1]{\begin{eqnarray} \label{(#1)}}
\newcommand{\ea}{\end{eqnarray}}
\newcommand{\nn}{\nonumber}
\newcommand{\rf}[1]{(\ref{(#1)})}

\def\rp{$R_p \hspace{-1em}/\;\:$}
\def\rpm{R_p \hspace{-0.8em}/\;\:}

\def\pmb#1{\setbox0=\hbox{#1}%
  \kern-.015em\copy0\kern-\wd0
  \kern.03em\copy0\kern-\wd0
  \kern-.015em\raise.0233em\box0 }
\def \znbb {0\nu\beta\beta}

\def\bfr{\pmb{${r}$}}

\begin{document}

\begin{center}

{\bf On the SUSY Accompanied 
Neutrino Exchange Mechanism of
Neutrinoless Double Beta Decay}
\bigskip

{M. Hirsch, H.V. Klapdor-Kleingrothaus and S.G. Kovalenko$^*$
\bigskip

{\it
Max-Planck-Institut f\"{u}r Kernphysik, P.O. 10 39 80, D-69029,
Heidelberg, Germany}

$^*${\it Joint Institute for Nuclear Research, Dubna, Russia}
}

\end{center}

\begin{abstract}

The neutrinoless double beta decay ($\znbb$) induced by light 
Majorana neutrino exchange between decaying nucleons, accompanied
by the squark exchange inside one nucleon, recently discussed by 
Babu and Mohapatra, is carefully analyzed both from the particle 
and nuclear physics sides. New nuclear matrix elements relevant 
to this mechanism are calculated. We extend the analysis to include 
mixing of light neutrinos with heavy and "sterile" neutrinos. It 
introduces  another supersymmetric (SUSY) contribution to $\znbb$.
We discuss constraints on the \rp MSSM parameters imposed by   
the current experimental limit on $\znbb$ decay half-life 
of $^{76}$Ge. 
\end{abstract}
%%%%%%%%%%%%%%%%%%%%%%%%%%%%%%%%%%%%%%%%%%%%%%%%%%%%%%%%%%%%%%%%%%%
%
%\section{Introduction} 
%
Neutrinoless double beta ($\znbb$) decay is a sensitive probe of 
physics beyond the standard model, since it violates lepton number. 
Recent experimental progress has pushed the existing half-life 
limits of $\znbb$ decay beyond $T_{1/2}(\znbb)$ $\ge 7.4 \times 10^{24}$ 
years and further progress can be expected in the near future 
\cite{hdmo94}. This experimental result casts stringent constraints 
on new physics. Particularly, for the conventional mechanism of 
$\znbb$-decay with massive Majorana neutrino exchange between 
decaying nucleons (see fig. 1) it implies an upper bound on the 
neutrino mass below $1$ eV \cite{hdmo94}. (There exist, however, 
other mechanisms 
which might induce $\znbb$ decays as well \cite{doi85}, \cite{dbd_mech}.)

In this paper we study contributions to $\znbb$ decay within the 
R-parity violating Minimal Supersymmetric Standard Model (\rp  MSSM). 
The \rp  MSSM has been extensively discussed in the literature 
since it has  very interesting phenomenological \cite{phenomen} 
and cosmological \cite{Cosmology} implications. It also  gives a 
very natural framework for rare lepton number violating processes 
and particularly $\znbb$ decay \cite{Mohapatra}-\cite{BM}.

The supersymmetric mechanism of $\znbb$ decay was first proposed
by Mohapatra \cite{Mohapatra} and later studied in more details in 
Refs. \cite{Vergados}, \cite{HKK1}. In Ref. \cite{HKK2} it was 
shown that the gluino exchange contribution to $\znbb$-decay leads 
to a very stringent limit on the first generation \rp Yukawa 
coupling $\lambda'_{111} \leq 3.9\cdot 10^{-4}$. Recently, 
Babu and Mohapatra \cite{BM} found another contribution comparable 
in size with the gluino exchange.  It allows one to set stringent 
limits on combinations of the intergeneration \rp-Yukawa couplings 
such as $\lambda'_{11i}\lambda'_{1i1}$, where $i$ denotes generations.

However, in ref. \cite{BM} limits were deduced using a simplified 
estimation of the nuclear structure matrix elements. The approach 
used in \cite{BM} is based on a simple replacement of the virtual 
particle momenta and energies by the (estimated) Fermi momentum 
$p_F$ and energy $E_F$ of a nucleon inside the nucleus, therefore 
neglecting essentially all nuclear structure effects. Since the 
stringent constraints found in \cite{BM} may have important
consequences for the \rp MSSM  phenomenology, it is desirable to 
substantiate these results by detailed calculations.

In this letter we present the results of detailed particle and nuclear
physics calculations relevant to the contribution of the diagrams 
in fig. 2 to $\znbb$ decay. 
We extend the previous work to include mixing of the standard light 
neutrinos with exotic heavy neutrinos, neutralinos or some other 
neutral heavy particles or light "sterile" neutrino singlets 
\cite{sterile}. Inclusion of mixing leads to the new slepton 
exchange diagram in fig. (2.b). 

%
%\section{Formalism}
%

The \rp MSSM  is an extension of the MSSM which results from the 
inclusion of explicit $R$-parity violating (\rp) terms  $W_{\rpm}$ 
into the superpotential $ W = W_{MSSM} + W_{\rpm} $, where  

\ba{superpotential}
W_{\rpm} &=& \lambda_{ijk}L_i L_j {\bar E}_k 
           + \lambda'_{ijk}L_i Q_j {\bar D}_k
           + \lambda''_{ijk}{\bar U}_i {\bar D}_j {\bar D}_k
\ea
Indices $i,j, k$ stand for generations. $L$,  $Q$ denote lepton 
and quark doublet superfields and ${\bar E}, \ {\bar U},\  {\bar D}$ 
lepton and {\em up}, {\em down} quark singlet  superfields.
The first two terms in eq. \rf{superpotential} lead to lepton number 
violation, while the last one violates baryon number. For 
$\znbb$ decay only the $\lambda$ and $\lambda'$ type 
couplings are of relevance. 

Mixing between scalar superpartners $\tilde f_{L,R}$ of the left 
and right-handed fermions $f_{L,R}$ will play a crucial 
role in our subsequent consideration. It occurs due to non-diagonality 
of the mass matrix which can be written as 
\ba{e4}
{\cal M}^2_{\tilde f} &= \mbox{$ \left( \begin{array}{cc}
m^2_{\tilde{f}_L} + m_f^2 - 0.42 D_Z & - m_f (A_f + \mu \tan\! \beta) \\
- m_f (A_f + \mu \tan\! \beta) & m^2_{\tilde{f}_R} + m_f^2 - 0.08 D_Z
\end{array} \right) $}. 
\ea
Here,  $ f = d,  s,  b,  e,  \mu,  \tau$  and $\tilde f$ are 
their superpartners. $D_Z = M_Z^2 \cos\! 2\beta$ with 
$\tan\!\beta =\langle H_2^0\rangle / \langle H_{1}^{0} \rangle$ 
being the ratio of vacuum 
expectation values of the two Higgs doublets, $m_{\tilde f_{L,R}}$ 
are soft sfermion masses, $A_f$  are soft SUSY breaking parameters 
describing the strength of trilinear scalar interactions, and $\mu$ 
is the supersymmetric Higgs(ino) mass parameter. Once sfermion  
mixing is included, the current eigenstates $\tilde{f}_L, \tilde{f}_R$
become superpositions of the mass eigenstates $\tilde{f}_i$ with 
the masses $m_{\tilde f_i}$ and the corresponding  mixing angle 
$\theta^f$ is defined as
\ba{mixing}\nn
m^2_{\tilde q_{1,2}}=\frac{1}{2}
\left[m^2_{LL} + m^2_{RR}\mp \sqrt{(m^2_{LL}-m^2_{RR})^2 + 4 m^4_{LR}}
\ \right]; \\
\label{(mixangle)}
\sin\! 2\theta^f = \frac{2 m^2_{(f)LR}}
{m^2_{\tilde{f}_1}-m^2_{\tilde{f}_2}},
\ea
where $m^2_{LR}, m^2_{LL}, m^2_{RR}$ denote the $(1,2), (1,1), (2,2)$ 
entries of the mass matrix (\rf{e4}). 

Now it is straightforward to find the  effective 4-fermion 
$\nu-u-d-e$ vertex induced by the sfermion exchange in the 
diagrams presented in fig. 2. The corresponding effective 
Lagrangian, after a Fiertz rearrangement, takes the form
\ba{L_SUSY}\nn
{\cal L}_{SUSY}^{eff}(x) &=&
\frac{G_F}{\sqrt{2}} \left[ \frac{1}{4} 
\left(\eta_{(q)LR}^{nj}  - 4 \eta_{(l)LR}^{nj} \right)\cdot U^{*}_{ni} 
\cdot \left(\bar \nu_i (1 + \gamma_5) e^c_j\right) 
\left(\bar u (1 + \gamma_5)  d\right)   - \right. \\ 
&-& 2 \eta_{(l)LL}^{nj} \cdot U_{ni}  
\cdot \left(\bar \nu_i (1 - \gamma_5) e^c_j\right) 
\left(\bar u (1 + \gamma_5)  d\right) + \\ \nn
&+& \frac{1}{2}\eta_{(q)RR}^{nj}\cdot U_{ni} 
\left(\bar \nu_i\ \gamma^{\mu} (1 + \gamma_5)  e^c_j\right) 
\left(\bar u\   \gamma_{\mu} (1 - \gamma_5)  d\right)  + \\ \nn
&+& \left. \frac{1}{8}\eta_{(q)LR}^{nj}\cdot U^{*}_{ni}  \cdot 
\left(\bar \nu_i\ \sigma^{\mu\nu}  (1 + \gamma_5)  e^c_j\right) 
\left(\bar u\   \sigma_{\mu\nu} (1 + \gamma_5)  d\right)\right].
\ea 
The \rp MSSM  parameters $\eta$ and neutrino mixing matrix 
$U_{ij}$ are defined as follows
\ba{eta}
%%%%%%%%%%%%%%%%%%%%%%%%%%%%%%%%%%%%%%%%%%%%%%%%%%%%%%%%
\eta_{(q)LR}^{nj} &=& \sum_{k} \frac{\lambda'_{j1k}\lambda'_{nk1}}{2
\sqrt{2} G_F} 
\sin{2\theta^{d}_{(k)} }\left( \frac{1}{m^2_{\tilde d_1 (k)}} -  
\frac{1}{m^2_{\tilde d_2 (k)}}\right), \\
%%%%%%%%%%%%%%%%%%%%%%%%%%%%%%%%%%%%%%%%%%%%%%%%%%%%%%%%
\eta_{(q)RR}^{nj} 
&=& \sum_{k} \frac{\lambda'_{j1k}\lambda'_{n1k}}{2 \sqrt{2} G_F}
\left( \frac{\sin{\theta^{d}_{(k)} }}{m^2_{\tilde d_1 (k)}} +  
\frac{\cos\theta^{d}_{(k)}}{m^2_{\tilde d_2 (k)}}\right), \\
%%%%%%%%%%%%%%%%%%%%%%%%%%%%%%%%%%%%%%%%%%%%%%%%%%%%%%%%
\eta_{(l)LR}^{nj} &=&\sum_{k} \frac{\lambda'_{k11}\lambda_{njk}}
 {2 \sqrt{2} G_F}
 \sin{2\theta^{e}_{(k)} }\left( \frac{1}{m^2_{\tilde e_1 (k)}} -  
\frac{1}{m^2_{\tilde e_2 (k)}}\right), \\
%%%%%%%%%%%%%%%%%%%%%%%%%%%%%%%%%%%%%%%%%%%%%%%%%%%%%%%%
\eta_{(l)LL}^{nj}&=&  \sum_{k} \frac{\lambda'_{k11}\lambda_{nkj}}
 {2 \sqrt{2} G_F}
\left(\frac{\cos{\theta^{e}_{(k)}}}{m^2_{\tilde e_1 (k)}} +  
\frac{\sin\theta^{e}_{(k)}}{m^2_{\tilde e_2 (k)}}\right),\\
%%%%%%%%%%%%%%%%%%%%%%%%%%%%%%%%%%%%%%%%%%%%%%%%%%%%%%%%
\label{(neutr1)}
\nu^0_i &=& \sum_{j} U_{ij} \nu_j.
%%%%%%%%%%%%%%%%%%%%%%%%%%%%%%%%%%%%%%%%%%%%%%%%%%%%%%%%
\ea
Here $\eta_{(f)LR}$ denotes the contribution vanishing in the 
absence of $\tilde f_L-\tilde f_R$ - mixing while $\eta_{(f)LL}$ 
and $\eta_{(f)RR}$ in this limit correspond to the $\tilde f_L$ and 
$\tilde f_R$ exchange contribution in fig. 2. We use the notations 
$d_{(k)} = d, s, b$ and $e_{(k)} = e, \mu, \tau$. Due to the antisymmetry 
of the Yukawa coupling $\lambda_{njk}$ in $nj$ it follows that 
$\eta_{(l)LR}^{nn} = 0$. This is an essential difference between 
the slepton $\tilde l_L-\tilde l_R$ and the squark 
$\tilde q_L-\tilde q_R$ contributions. The latter is not imposed 
to vanish at any combination of indexes.     

The matrix element of the SUSY accompanied neutrino exchange 
mechanism can then be calculated according to the 
the diagrams in fig. 2 with the point-like 4-fermion vertex 
described by the effective Lagrangian eq. \rf{L_SUSY} with 
the sfermion exchange parts in the top.  The bottom parts of these 
diagrams is the standard model charged current 
(SMCC) interaction. Applying the standard procedure (for details see 
\cite{HKK2}), one can get the matrix element 
${\cal R}_{\znbb}(0^+\rightarrow 0^+)$ of the $\znbb$ decay for 
$0^+\rightarrow 0^+$ transitions. For two outgoing electrons 
in S-wave states it takes the form
\ba{R_0nu}
{\cal R}_{\znbb}(0^+\rightarrow 0^+) &=& C_{0\nu} f^2_A 
\left[a\cdot \bar e (1 + \gamma_5) e^c 
+  b\cdot \bar e \gamma_0\gamma_5 e^c\right],\\ 
%%%%%%%%%%%%%%%%%%%%%%%%%%%%%%%%%%%%%%%%%%%%%%%%%%%%%%%%%
\label{(a)}
a &=&  \left(4 \eta_{(l)LR}^{n1} - \eta_{(q)LR}^{n1}\right)
 U^{*}_{ni} U_{ei} {\cal M}^{(i)}_1 \left(m_e R\right)^{-1}+\\ \nn
%%%%%%%%%%%%%%%%%%%%%%%%%%%%%%%%%%%%%%%%%%%%%%%%%%%%%%%%%%
&+& \left(\eta_{(q)RR}^{n1} - \delta_{ne}\right) U_{ni} U_{ei}
 \frac{m_{\nu_i}}{m_e} 
 {\cal M}^{(i)}_2, \\
%%%%%%%%%%%%%%%%%%%%%%%%%%%%%%%%%%%%%%%%%%%%%%%%%%%%%%%%%
\label{(c)}
b &=& 4\  \eta_{(l)LL}^{n1} U_{ni} U_{ei}  \frac{m_{\nu_i}}{m_e} 
{\cal M}^{(i)}_3.
\ea
The term proportional to $\delta_{ne}$ corresponds to the ordinary 
neutrino exchange contribution with two standard model charged current
vertices in fig. 1. The normalization factor $C_{0\nu}$ is defined 
as $ C_{0\nu} = (G_F^2 2 m_e)/(8\sqrt{2}\pi R)$.
We would like to stress that the terms proportional to $b$ in 
eq. \rf{R_0nu} can, in principle, be discriminated from the terms 
proportional to $a$ (particularly from the ordinary mass mechanism 
of $\znbb$-decay) by measuring the angular correlation between 
the two outgoing electrons. 

The following nuclear matrix elements are involved in the calculation 
of $R_{\znbb}$ in eq. \rf{R_0nu}
\ba{Omega}
{\cal M}^{(i)}_{1} &=& \alpha_1 \left[{\cal M}^{(i)}_{T'} + 
\frac{1}{3} {\cal M}^{(i)}_{GT'}\right], \\ 
%%%%%%%%%%%%%%%%%%%%%%%%%%%%%%%%%%%%%%%%%%%%%%%%%%%%%%%%
{\cal M}^{(i)}_{2} &=& \alpha_2 {\cal M}^{(i)}_{F}  -
{\cal M}^{(i)}_{GT}, \ \ 
%%%%%%%%%%%%%%%%%%%%%%%%%%%%%%%%%%%%%%%%%%%%%%%%%%%%%%%%
{\cal M}^{(i)}_{3} = \alpha_3 {\cal M}^{(i)}_{F}.
%%%%%%%%%%%%%%%%%%%%%%%%%%%%%%%%%%%%%%%%%%%%%%%%%%%%%%%%
\ea
They are defined by (summation over nucleons is suppressed) 
\ba{nucl_matr}
%%%%%%%%%%%%%%%%%%%%%%%%%%%%%%%%%%%%%%%%%%%%%%%%%%%%%%%%
{\cal M}^{(i)}_{F} &=& <0^+_f|| h_{+}(r_{ab}, m_{\nu_i})
                            \tau^+_a\tau^+_b||0^+_i>,\\ 
%%%%%%%%%%%%%%%%%%%%%%%%%%%%%%%%%%%%%%%%%%%%%%%%%%%%%%%%
{\cal M}^{(i)}_{GT} &=& <0^+_f|| h_{+}(r_{ab}, m_{\nu_i})
                     (\vec \sigma_a\vec\sigma_b)\tau^+_a\tau^+_b||0^+_i>, \\ 
%%%%%%%%%%%%%%%%%%%%%%%%%%%%%%%%%%%%%%%%%%%%%%%%%%%%%%%%
{\cal M}^{(i)}_{GT'} &=& <0^+_f|| h_R(r_{ab}, m_{\nu_i})
                     (\vec \sigma_a\vec\sigma_b)\tau^+_a\tau^+_b||0^+_i>, \\ 
%%%%%%%%%%%%%%%%%%%%%%%%%%%%%%%%%%%%%%%%%%%%%%%%%%%%%%%%
{\cal M}^{(i)}_{T'} &=& <0^+_f||h_{T'}(r_{ab}, m_{\nu_i})\left[
(\vec \sigma_a {\hat{\bf r}_{ab}^{~}})(\vec \sigma_b {\hat{\bf r}_{ab}^{~}}) -
\frac{1}{3}(\vec \sigma_a\vec\sigma_b) \right]\tau^+_a\tau^+_b||0^+_i>. 
%%%%%%%%%%%%%%%%%%%%%%%%%%%%%%%%%%%%%%%%%%%%%%%%%%%%%%%%%
\ea
Neutrino potentials can be written in the integral form
\ba{potent}
h_{+}(r_{ab}, m_{\nu_i}) &=& \frac{2}{\pi} R \int_0^{\infty}
dq\cdot q^2 \frac{j_0(q r_{ab}) f^2(q^2)}{\omega(\omega + \bar A)}, \\ 
%%%%%%%%%%%%%%%%%%%%%%%%%%%%%%%%%%%%%%%%%%%%%%%%%%%%%%%%%
h_{R}(r_{ab}, m_{\nu_i}) &=& \frac{2}{\pi} \frac{R^2}{m_P} \int_0^{\infty}
dq\cdot q^4 \frac{j_0(q r_{ab}) f^2(q^2)}{\omega(\omega + \bar A)}, \\ 
%%%%%%%%%%%%%%%%%%%%%%%%%%%%%%%%%%%%%%%%%%%%%%%%%%%%%%%%%
h_{T'}(r_{ab}, m_{\nu_i}) &=& \frac{2}{\pi} \frac{R^2}{m_P} \int_0^{\infty}
dq\cdot q^4 \frac{j_0(q r_{ab}) - 3 j_1(q r_{ab})}{\omega(\omega + \bar A)} 
f^2(q^2).
%%%%%%%%%%%%%%%%%%%%%%%%%%%%%%%%%%%%%%%%%%%%%%%%%%%%%%%%%
\ea
Here, $\omega = \sqrt{q^2 + m_{\nu_i}^2}$; $j_k(q r)$ are 
spherical Bessel functions and $R_0$ is the nuclear radius, 
introduced to make the matrix elements dimensionless. 
The following notations are used: 
${\bf r}_{ab} = ({\overrightarrow r}_a - {\overrightarrow r}_b ),\
r_{ab} = |{\bfr}_{ab}|, \  {\hat{\bf r}_{ab}^{~}}= {\bf r}_{ab}/r_{ab}$. 
The above formulae have been written in the closure approximation
which is well motivated for the $\znbb$ decay ~\cite{doi85}. 
${\bar A}$ in eq. \rf{potent} is the average intermediate state energy. 

Note that the nuclear matrix element ${\cal M}^{(i)}_{T'}$ has never 
been considered in the literature before. We have calculated the 
nuclear matrix elements relevant for our subsequent numerical 
analysis using the pn-QRPA model of \cite{mut89}. We take
into account both short-range correlations and finite nucleon 
size effects. The latter is described by introducing nucleon form 
factors in momentum space. In the present case these are the SMCC 
form factors $F_{(V,A)}^{CC}(q^2)$ and the isovector scalar and 
pseudoscalar current form factors $F^3_{P,S}(q^2)$.
For all form factors we take, as usual, a dipole parameterization: 
$F^a_{i}(q^2)/F^a_{i}(0) =  f(q^2) = \left(1 + q^2/m_A^2 \right)^{-2}$
with $m_A = 0.85$GeV. The q-dependent factor $f(q^2)$ is included in 
the definition of the neutrino potentials in eqs. \rf{potent}. 
The form factor normalizations  are 
$F^{CC}_{V(A)}(0) = f_{V(A)}\approx 1(1.261)$. $F^{3}_{S,P}(0)$ 
can be calculated within the conventional non-relativistic quark 
model or the bag model. We take their numerical values from ref. 
\cite{Adler} $F^{3}_{S}(0) \approx 0.48$ $F^{3}_{P}(0) \approx 4.41$. 
These values correspond to the bag model calculations.

The nucleon structure coefficients in \rf{Omega} are defined as 
$\alpha_1 = (F^{(3)}_P/(2 f_A)$, $\alpha_2 = (f_V/f_A)^2$,   
$\alpha_3 = (f_V/f_A)(F^{(3)}_S/f_A)$.

Now having the $\znbb$ matrix element completely specified 
the inverse half life $T_{1/2}(\znbb)$ can be written as
\ba{half-life}
T^{-1}_{1/2}(\znbb) = |a|^2 G_{01} + |b|^2 h_9 G_{09} +  Re(a^* b) h_6 G_{06},
\ea
where $ h_6 = m_e R/8, h_9 = (m_e R)^2/16$, and phase space factors $G_{0i}$
are given in ref. \cite{doi85}. 

The formulae presented above describe the contribution of the 
diagrams in fig. 2 for a general neutrino content (see eq. \rf{neutr1}). 
We now turn to a particular case and assume that all neutrino mass 
eigenstates fall into two groups: light neutrinos $m_{\nu_i} < 10$MeV 
and heavy neutrinos $m_{\nu_i} > 10$GeV. We denote them as $\nu_i$ 
and $N_i$, respectively. There might be also a ``sterile'' neutrino 
$\nu_s$ with respect to the SM gauge group. Assume further, that the 
light neutrinos have non-negligible mixing with some heavy neutrinos 
$N_i$ or probably with the ``sterile'' neutrino $\nu_s$. Than the 
neutrino composition \rf{neutr1} and the unitarity relation for the 
mixing matrix $U_{ij}$ can be written as
\ba{neutr2}
\nu^0_i &=& \sum'_{j} U_{ij} \nu_j + \sum''_{j} U_{ij} \nu_j +  U_{is} \nu_s,\\
\sum_{j} U^*_{ij} U_{ei} &=& \sum'_{j} U^*_{ij} U_{ei} + \Delta_i, \ \
\Delta_i = \sum''_{j} U^*_{ij} U_{ei} +   U_{is} \nu_s.
\ea
Such a structure of the neutrino sector breaks contributions of each
nuclear matrix elements in eq. \rf{R_0nu} into two pieces with a simple
dependence on the neutrino mass. As an example of this effect consider 
the first term in eq. \rf{a}
\ba{breaking}
\sum_i U^{*}_{ni} U_{ei} {\cal M}^{(i)}_1 = 
\sum'_i U^{*}_{ni}U_{ei}{\cal M}^{\nu}_1  
+ \sum''_i \frac{U^{*}_{ni} U_{ei}}{M^2_{N_i}} {\cal M}^{N}_1.
\ea
The "sterile" neutrino does not contribute being a SM singlet.
The matrix elements are 
${\cal M}^{\nu}_1 = {\cal M}^{(i)}_1(m_{\nu}=0)$ 
and ${\cal M}^{N}_1 = \lim_{M_N^2\rightarrow\infty} M_N
{\cal M}^{(i)}_1(M_N)$. They do not depend on neutrino masses. 
Proceeding in a similar way with
the other matrix elements in eq. \rf{R_0nu} and substituting the
result in the half-life formula eq. \rf{half-life} we obtain a
polynomial in $m_{\nu}$ and $M_N^{-1}$. Assume light neutrinos $\nu$
to be very light while heavy neutrinos $N$ to be very heavy so that
one can neglect terms depending on these masses. Than, keeping only 
leading terms we retain 
\ba{dominant}
T^{-1}_{1/2}(\znbb) = G_{01} {\cal M}^{\nu}_1 \left(m_e R\right)^{-1}
\left(4 \bar\eta_{(l)} - \bar\eta_{(q)}  + \eta_{(q)}\right)^2,
\ea
where eq. \rf{breaking} and the property $\eta_{(l)LR}^{nn} = 0$ 
have been used. We denoted $\eta_{(q)} = \eta_{(q)LR}^{11}$
and introduced the effective parameters as 
\ba{efpar}
\bar\eta_{(l,q)} = \sum_n\Delta_n \eta_{(l,q)LR}^{n1}
\ea
For the $\bar\eta_{(l)}$ summation starts from $n=2$.
The nuclear matrix element ${\cal M}^{\nu}_1$ in eq. \rf{dominant} 
can be directly obtained from eq. \rf{Omega} for ${\cal M}^{(i)}_1$ 
as explained after eq. \rf{breaking}. This matrix element has never 
been calculated in the literature before. Its value calculated in
the pn-QRPA for the particular case of $^{76}$Ge analyzed below 
is ${\cal M}^{(\nu)}_1({}^{76}\mbox{Ge}) = 2.1$. \footnote{Recall 
that this numerical value corresponds to our dimensionless 
convention.}

Now we are ready to discuss constraints on the parameters in eq. 
\rf{dominant} imposed by the current experimental lower half-life 
limit.  We use the result from the Heidelberg-Moscow
$^{76}$Ge experiment 
  $T_{1/2}^{\znbb}(^{76}Ge, 0^+ \rightarrow 0^+) > 7.4 \times 10^{24}$ 
$years \  90\% \ c.l.$

Disregarding a situation with unnatural fine-tuning between the 
different terms in eq. \rf{dominant} one can extract individual 
limits, numerically $\eta_{(q)} \leq 2.9 \times 10^{-8}$,  
$\bar\eta_{(l)}  \leq 7.2 \times 10^{-9}$. 
These constraints lead, in principle, to a multidimensional exclusion
curve. A simplified picture can be obtained under some reasonable 
assumptions. Assume all the MSSM mass parameters in eqs. 
\rf{eta} and \rf{e4}, \rf{mixangle} to be approximately equal to 
the "effective" SUSY breaking scale $\Lambda_{SUSY}$.
Than we get a simplified set of constraints
\ba{limit3}
\lambda'_{11i}\lambda'_{1i1} \leq \epsilon'_i 
\left(\frac{\Lambda_{SUSY}}{100 GeV}\right)^3,\ \ 
\Delta_n\lambda'_{311}\lambda_{n13}  \leq  \epsilon
\left(\frac{\Lambda_{SUSY}}{100 GeV}\right)^3.
\ea
In the last equation we kept only the term corresponding to 
$\tilde\tau$ exchange. For the known values of SM quark 
and lepton masses in the following we use $m_{d_1} = m_{d} = 7.5$MeV, 
$m_{d_2} = m_{s} = 150$MeV, $m_{d_3} = m_{b} = 4.5$GeV  and  
$m_{\tau} = 1.8$GeV. These masses are present in eqs. 
\rf{e4}, \rf{mixangle}. Then, 
$\epsilon'_{1,2,3} =\{6.4\times 10^{-5},
3.2\times 10^{-6},1.1\times 10^{-7}\}$ and 
$\epsilon = 6.5\times 10^{-8}$. 

To obtain information on the \rp Yukawa couplings themselves 
one may use some reasonable values for $\Lambda_{SUSY}$.
If one takes, following ref. \cite{BM}, the value  
$\Lambda_{SUSY} = 100$ GeV lying in the region of the current
experimental lower bound for the SUSY particles, one gets 
$ \lambda'_{113}\lambda'_{131}\leq 1.1 \times 10^{-7} $,
$ \lambda'_{112}\lambda'_{121}\leq 3.2 \times 10^{-6} $, 
$ {\lambda'}^2_{111} \leq 6.4 \times 10^{-5}$
and 
$\Delta_n\lambda'\lambda \leq 6.5 \times 10^{-8}$. The former two
limits correspond to those in ref. \cite{BM}, but are somewhat 
weaker than quoted by Babu and Mohapatra. This difference 
can be partly traced back to our numerical value of the 
matrix element, which turns out to be smaller than anticipated. 
The limit on ${\lambda'}^2_{111}$ can be compared with the 
corresponding limit obtained from the gluino exchange diagram 
in ref. \cite{HKK2}. The latter being 
${\lambda'}^2_{111} \leq  1.52\cdot10^{-7}$ is much more 
stringent then the one derived here. 
More conservative estimations can be derived from eq. \rf{limit3} 
implying $\Lambda_{SUSY}\sim 1$TeV motivated by the SUSY naturalness 
arguments. 
Then,  
$ \lambda'_{113}\lambda'_{131}\leq 1.1 \times 10^{-4} $,
$ \lambda'_{112}\lambda'_{121}\leq 3.2 \times 10^{-3} $, 
$ {\lambda'}^2_{111} \leq 6.4 \times 10^{-2}$
and 
$\Delta_n\lambda'\lambda \leq 6.5 \times 10^{-4}$.

Neglecting mixing between the light SM non-singlet neutrinos and the
non-standard sector discussed above, one arrives at the case
considered by Babu and Mohapatra \cite{BM}. In this case $\Delta_n = 0$ 
and in turn $\bar\eta{(l,q)} = 0$. Then the squark exchange diagram in
fig. 2(a) is the only contribution  to eq. \rf{dominant}. Introducing 
mixing one can obtain new information about $\lambda$ type 
interactions. In this case, however, this information is accessible 
only in a form of 
some effective value of the \rp Yukawa couplings 
$\Delta\lambda'\lambda$. The upper limit  
for the couplings itself implies an extra uncertainty 
due to the unknown mixing factor $\Delta$.

The new contributions analysed in this letter together with the 
previously discussed in the literature \cite{Mohapatra}-\cite{HKK2} 
complete the tree-level mechanisms of  $\znbb$ decay within 
the R-parity violating Minimal Supersymmetric Standard Model 
(\rp  MSSM). A detailed presentation of the analysis will be 
given elsewhere. 
 
In conclusion we would like to stress that the current 
experimental limit on $\znbb$ decay half life allows one to 
establish rather stringent limits on \rp violating SUSY interactions.
This might be an additional valuable motivation for present 
and forthcoming $\znbb$ decay experiments.

\bigskip
\centerline{\bf ACKNOWLEDGMENTS}

We thank V.A. Bednyakov and  R.N. Mohapatra for helpful discussions.
The research described in this publication was made possible 
in part (S.G.K.) by Grant GNTP 215NUCLON from the Russian ministry 
of science. M.H. would like to thank the Deutsche 
Forschungsgemeinschaft for financial support by grants kl 253/8-1 and 
446 JAP-113/101/0.

%\newpage

{\large\bf Figure Captions}\\

\begin{itemize}

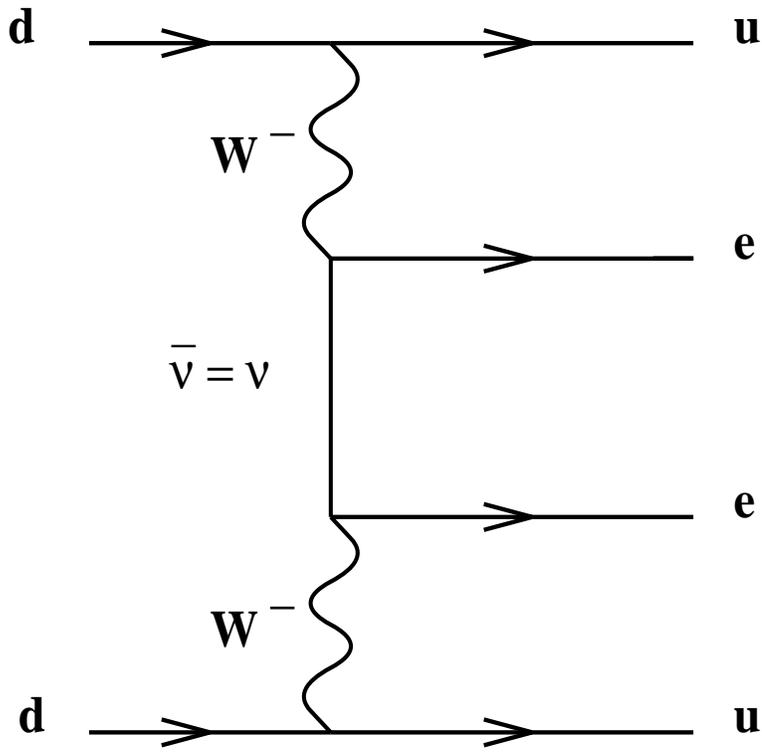
\item[Fig.1] Feynman graphs for the conventional mechanism
of $\znbb$ decay by exchange of a massive Majorana neutrino.

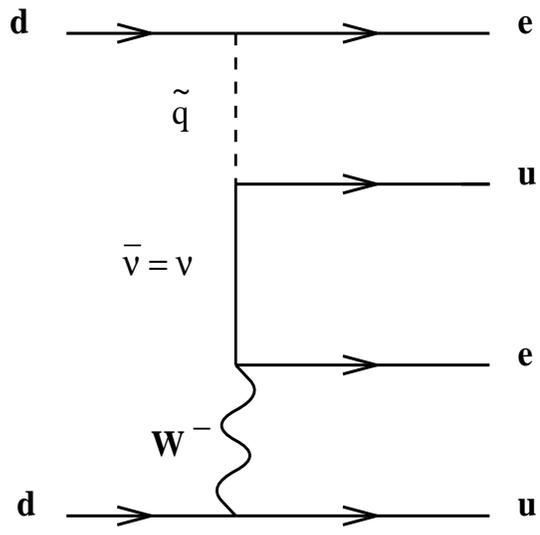
\item[Fig.2]  Feynman graphs for the supersymmetry accompanied Majorana
              neutrino exchange mechanism of the $\znbb$ decay corresponding 
              to (a) squarks and 
                 (b) sleptons contributions.

\end{itemize}

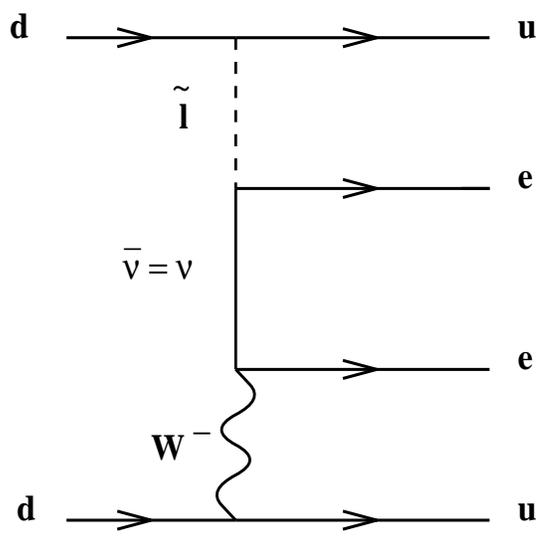
\end{document}